\documentclass[twocolumn,showpacs,preprintnumbers,showkeys,superscriptaddress]{revtex4}
\usepackage{graphicx}
\usepackage{dcolumn}
\usepackage{bm}

\def\eqref#1{Eq.~(\ref{eq:#1})}

\begin{document}

\title {Number of states for nucleons in a single-$j$ shell}
\author{Y. M. Zhao}     \email{ymzhao@sjtu.edu.cn}
\affiliation{Department of Physics,  Shanghai Jiao Tong
University, Shanghai 200240, China}  \affiliation{Cyclotron
Center, Institute of Physical Chemical Research (RIKEN), Hirosawa
2-1, Wako-shi,  Saitama 351-0198,  Japan}\affiliation{Center of
Theoretical Nuclear Physics, National Laboratory of Heavy Ion
Accelerator, Lanzhou 730000, China}
\author{A. Arima} \affiliation{Science Museum,
Japan Science Foundation, 2-1 Kitanomaru-koen, Chiyodaku, Tokyo
102-0091, Japan}

\date{\today}

\begin{abstract}
In this paper we obtain number of states with a given  spin $I$
and a given isospin $T$ for systems with three and four nucleons
in a single-$j$ orbit, by using sum rules of six-$j$ and nine-$j$
symbols obtained in earlier works.
\end{abstract}


\maketitle

Recently, there have been efforts to obtain algebraic formulas for
the number of spin $I$ states (denoted by $D_I$ in this paper) for
fermions in a single-$j$ shell (We use a convention that $j$ is a
half integer) [1-5]. So far most discussions have been restricted
to identical particles. In nuclear physics, there are two types of
valence nucleons, i.e., protons and neutrons. Therefore, it is
also interesting to obtain the formulas of number of states with a
given spin $I$  and isospin $T$  (denoted by $D_{IT}$ in this
paper) which includes automatically $D_I$ for identical particles
because $D_I$ for identical particles studied in earlier works
equals $D_{IT}$ with $T=T_{\rm max}$.

In Refs. \cite{Zhao-prc70} and \cite{sum-rule} sum rules of
six-$j$ and nine-$j$ symbols were studied by using the summation
(trace) of diagonal matrix elements of individual $J$ pairing
interaction for three and four identical particles in a single-$j$
shell. If one takes all two-body matrix elements to be 1, i.e.,
strength $G_J$ of all $J$ pairing interaction equals 1, the
summation of traces over $J$ must equal $\frac{n(n-1)}{2}$
multiplied by the number of spin $I$ states. This is nothing but
the trace of identity 1.

In this paper we shall go in a reversed direction:  we obtain
formulas of   $D_{IT}$  by using the sum rules of six-$j$ and
nine-$j$ symbols obtained in Refs. \cite{Zhao-prc70} and
\cite{sum-rule}. Similarly to Ref.\cite{sum-rule}, we first define
the $J$-pairing interaction $H_{JT}$ for nucleons in a single-$j$
shell as follows.
\begin{eqnarray}
&&  H_{JT} = G_{JT} \sum_{M = -J}^J A_{MM_T}^{(JT) \dagger} \
A_{MM_T}^{(JT) }, \nonumber \\
&& A_{MM_T}^{(JT) \dagger} = \frac{1}{\sqrt{2}} \left[
a_{jt}^{\dagger}   a_{jt}^{\dagger}
     \right]^{(JT)}_{MM_T},
\nonumber \\
&&
     {A}_M^{(JT)} = - (-1)^{M+M_T} \frac{1}{\sqrt{2}} \left[
     \tilde{a}_{jt}
     \tilde{a}_{jt} \right]_{-M-M_T}^{(JT)}, ~ ~ ~
     \nonumber \\
&&
     \tilde{A}^{(JT)} = - \frac{1}{\sqrt{2}} \left[ \tilde{a}_{jt}
     \tilde{a}_{jt} \right]^{(JT)},  \label{H}
\end{eqnarray}
where $[~]_{MM_T}^{(JT)}$ means an operator in which two nucleons
are coupled to spin $J$ and isospin $T$  with spin projection $M$
and isospin projection $M_T$. We take $G_{JT}=1$ throughout this
paper.

In this paper we shall exemplify our method by using three and
four nucleons in a single-$j$ orbit. The same method can be
applied to to  three and four bosons with $F$ spin of interacting
boson model (IBM) II \cite{IBM}, IBM-III, and IBM-IV
\cite{IBM-III}.

First, one can prove that for $n=4$ the summation of all non-zero
eigenvalues of $H=H_{JT}$ is the trace of $H_{JT}$ matrix with
total spin $I$, and this trace is given by summing the diagonal
matrix elements
\begin{eqnarray}
&&  \langle 0 |\left[ A^{(JT_2)}   A^{(KT'_2)}
\right]^{(IT)}_{MM_T}    \left[ A^{(JT_2)\dag}
A^{(KT'_2)\dag} \right]^{(IT)}_{MM_T} |0 \rangle   \nonumber   \\
&=&1 + (-)^{I+T} \delta_{JK}
                    - 4 (2J+1) (2K+1) (2T_2+1) (2T'_2+1) \nonumber \\
&&  \times    \left\{%
     \begin{array}{ccc}
     j    & j  & J \\
     j    & j  & K \\
     J    & K  & I   \end{array}
     \right\} ~
     \left\{%
     \begin{array}{ccc}
     \frac{1}{2}    & \frac{1}{2}  & T_2 \\
     \frac{1}{2}    & \frac{1}{2}  & T'_2 \\
     T_2    & T'_2 & T   \end{array}
     \right\} ~
 \label{e5.12}
\end{eqnarray}
over $K$, $T_2$ and $T'_2$. Here  $T_2$ ($T'_2$) and $T$ are
isospins  for two and four nucleons, respectively.

The procedure to obtain   $D_{IT}$  is straightforward. From the
sum rule of two-body coefficients of fractional parentage, one
obtains $\frac{n(n-1)}{2}$ multiplied by  $D_{IT}$, if one sums
Eq. (\ref{e5.12}) over all allowed $J$, $K$, $T_2$ and $T'_2$;
namely,
\begin{eqnarray}
& & \sum_J \sum_{\alpha} \langle j^4  \alpha IT | H_J | j^4 \alpha
IT \rangle \nonumber \\
&=& \sum_{J K ~ T_2 T'_2}  \langle 0 |\left[ A^{(JT_2)}
A^{(KT'_2)} \right]^{(IT)}_{MM_T}  \left[ A^{(JT_2)\dag}
A^{(KT'_2)\dag}
\right]^{(IT)}_{MM_T} |0 \rangle  \nonumber \\
&  =&  6 D_{IT} ~. \label{sum-1}
\end{eqnarray}

For $n=4$, the maximum isospin $T$ ($T_{\rm max}$) should be equal
to 2.  For this case, $T_2$ and $T'_2$ in Eq.(\ref{sum-1}) equal
1,  $J$ and $K$ must take even values. Then we have
\begin{eqnarray}
& & 6 D_{I(T=2)} \nonumber \\
&=& \sum_{{\rm even} ~J {\rm even ~K}}  ( 1 + (-)^{I}
\delta_{JK} - 36 (2J+1) (2K+1)    \nonumber \\
&&  \times     \left\{
     \begin{array}{ccc}
     j    & j  & J \\
     j    & j  & K \\
     J    & K  & I   \end{array}
     \right\} ~
     \left\{
     \begin{array}{ccc}
     \frac{1}{2}    & \frac{1}{2}  & 1 \\
     \frac{1}{2}    & \frac{1}{2}  & 1 \\
     1    & 1 & 2   \end{array}
     \right\}   ~)~ \nonumber \\
&=&   \sum_{{\rm even} ~J ~ {\rm even ~K}}  ( 1 + (-)^{I}
\delta_{JK}     \nonumber \\
&& - 4 (2J+1) (2K+1) \left\{
     \begin{array}{ccc}
     j    & j  & J \\
     j    & j  & K \\
     J    & K  & I   \end{array}
     \right\}  ~ . \label{T=2}
\end{eqnarray}
The right hand side of above equation is just $6D_I$. This can be
easily understood by confirming the right hand side of
Eq.(\ref{T=2}) being equal to Eq.(8) of Ref. \cite{sum-rule}.
Although the formulas of $D_{IT=2}$  were available in Eqs. (3-5)
of Ref. \cite{Zhao}, we present $D_{IT=2}$ in new forms which are
simpler than other forms in practice. For $I \ge 2j-3$, the
formula of $D_{IT=2}$ in Ref. \cite{Zhao} is very simple and can
be easily applied. When $I\le 2j-3$, let us define $I=6k +
\kappa$, $L=\left[ \frac{j-\frac{6k+3}{2}}{3} \right]$, and $m=((j
- (6k+3)/2) ~ {\rm mod ~} 3) \equiv (j-3/2) ~ {\rm mod ~} 3$. We
have
\begin{eqnarray}
&& D_{I,T=2} = (3k+1) L + k m + 1 + 3 \left[\frac{k}{2} \right]
(\left[\frac{k}{2} \right] +1) \nonumber \\
&& + (k~{\rm mod ~} 2) (3\left[\frac{k}{2} \right] + 2)
\label{eq5}
\end{eqnarray}
when $\kappa = 0$;
\begin{eqnarray}
&& D_{I,T=2} = (3k+2) L + (k+1) m   \left[\frac{k}{2} \right]
(3\left[\frac{k}{2} \right] +2) \nonumber \\
&& + (k~{\rm mod ~} 2) (3\left[\frac{k}{2} \right] + 2)
\end{eqnarray}
when $\kappa = 2$; and
\begin{eqnarray}
&& D_{I,T=2} = (3k+3) L + (k+1) m +3 +  \left[\frac{k}{2} \right]
(3\left[\frac{k}{2} \right] +7) \nonumber \\
&& + (k~{\rm mod ~} 2) (3\left[\frac{k}{2} \right] + 4)
\end{eqnarray}
when $\kappa = 4$.

When $\kappa =4$, $D_{IT=2}$ can be further simplified. If
$I=12k+4$,
\begin{eqnarray}
&& D_{I,T=2} = (2k+1)j - \frac{1}{2} (34k+18k(k-1) + 3) \nonumber
\end{eqnarray}
and if $I=12k+10$,
\begin{eqnarray}
&& D_{I,T=2} = (2k+2)j - \frac{1}{2} (26k+9k(k-1) + 8) ~.
\nonumber
\end{eqnarray}

As in Ref. \cite{sum-rule}, we denote
\begin{eqnarray}
&& S_I (j^4, {\rm condition ~ X~ on }~ J {\rm ~and ~}K)
\nonumber \\
&& =  \sum_{{\rm X  }  } 4 (2J+1) (2K+1)
     \left\{
     \begin{array}{ccc}
     j    & j  & J \\
     j    & j  & K \\
     J    & K  & I   \end{array}
     \right\}   \label{condition}
\end{eqnarray}
for sake of simplicity.

Now we discuss the case of $T=1$. Here ($T_2$, $T'_2$) can take
following values: (1,0), (0,1), (1,1). Because of the Pauli
principle, there are requirements on  $J$ and $K$ values. The
corresponding requirements for ($J,K$) are: ($J={\rm even},K={\rm
odd}$), ($J={\rm odd},K={\rm even}$), and ($J={\rm even},K={\rm
even}$), respectively. We obtain
\begin{eqnarray}
& & 6 D_{I(T=1)} \nonumber \\
&=& \sum_{{\rm even} ~J {\rm even ~K}}  ( 1 - (-)^{I}
\delta_{JK} - 36 (2J+1) (2K+1)    \nonumber \\
&&  \times     \left\{
     \begin{array}{ccc}
     j    & j  & J \\
     j    & j  & K \\
     J    & K  & I   \end{array}
     \right\} ~
     \left\{
     \begin{array}{ccc}
     \frac{1}{2}    & \frac{1}{2}  & 1 \\
     \frac{1}{2}    & \frac{1}{2}  & 1 \\
     1    & 1 & 1   \end{array}
     \right\}   ~)~ \nonumber \\
&+& \sum_{{\rm odd} ~J {\rm even ~K}}  ( 1 - (-)^{I}
\delta_{JK} - 12 (2J+1) (2K+1)    \nonumber \\
&&  \times     \left\{
     \begin{array}{ccc}
     j    & j  & J \\
     j    & j  & K \\
     J    & K  & I   \end{array}
     \right\} ~
     \left\{
     \begin{array}{ccc}
     \frac{1}{2}    & \frac{1}{2}  & 1 \\
     \frac{1}{2}    & \frac{1}{2}  & 0 \\
     1    & 0 & 1   \end{array}
     \right\}   ~)~ \nonumber \\
&+& \sum_{{\rm even} ~J {\rm odd ~K}}  ( 1 - (-)^{I}
\delta_{JK} - 12 (2J+1) (2K+1)    \nonumber \\
&&  \times     \left\{
     \begin{array}{ccc}
     j    & j  & J \\
     j    & j  & K \\
     J    & K  & I   \end{array}
     \right\} ~
\left\{
     \begin{array}{ccc}
     \frac{1}{2}    & \frac{1}{2}  & 0 \\
     \frac{1}{2}    & \frac{1}{2}  & 1 \\
     0    & 1 & 1   \end{array}
     \right\}   ~)~ \nonumber \\
& = & \sum_{{\rm even} ~J ~ {\rm even ~K}}  ( 1 - (-)^{I}
\delta_{JK}) \nonumber  \\
&&+ 2\sum_{{\rm even} ~J ~ {\rm odd
~K}} ( 1 - (-)^{I}
\delta_{JK} )\nonumber \\
&-& \frac{1}{2} S(j^4, {\rm even} ~J ~ {\rm even ~K}) . \nonumber \\
\label{result1}
\end{eqnarray}
In derivation of Eq. (\ref{result1}) we used following relation:
\begin{eqnarray}
&&  \left\{   \begin{array}{ccc}
     \frac{1}{2}    & \frac{1}{2}   & 1 \\
     \frac{1}{2}    & \frac{1}{2}   & 1 \\
     1    & 1  & 1   \end{array}
     \right\}   = 0 ~,  \nonumber \\
&&   \left\{  \begin{array}{ccc}
     \frac{1}{2}     & \frac{1}{2}   & 0 \\
     \frac{1}{2}    & \frac{1}{2}   & 1 \\
     0    & 1  & 1   \end{array}
     \right\}   =  \left\{\begin{array}{ccc}
     \frac{1}{2}    & \frac{1}{2}  & 1 \\
     \frac{1}{2}    & \frac{1}{2}   & 0 \\
     1    & 0  & 1   \end{array}
     \right\}   = 1/6 ~.  \nonumber
\end{eqnarray}
To simplify Eq.(\ref{result1}), one should consider the number of
combinations of $J$ and $K$, as exemplified in Ref.
\cite{sum-rule}. We note that $J$, $K$ and $I$ must satisfy the
triangle relation for vector couplings.

When $n=4$ and $T=1$, $I_{\rm max}$ equals  $4j-3$ (odd value).
When $I\ge 2j$, let us define $I = I_{\rm max} - 2I_0$ for odd $I$
and $I_{\rm max} - 2I_0 -1$ for even $I$. Using Eq.(\ref{result1})
and Eq.(22) of Ref. \cite{sum-rule}, one can obtain
\begin{equation}
 D_{IT=1} = \left( \left[\frac{I_0}{2} \right] +1 \right)
 \left(\left[\frac{I_0}{2} \right]+1 + (I_0 {\rm ~mod~ } 2) \right)
 ~;
\end{equation}

When $I\le 2j$, we use Eqs. (\ref{result1}) and Eq.(21) of Ref.
\cite{sum-rule}, and obtain
\begin{eqnarray}
&&  D_{I(T=1)} = (I_0 + 1) j - \left( 1 + 4 \left[ \frac{I_0}{2}
 \right] + 6 \left[ \frac{I_0}{2}
 \right]^2  \right. \nonumber \\
 && ~~~~~~  + \left. (I_0 {\rm ~mod~} 2) (6\left[ \frac{I_0}{2}
 \right] + 3) \right) /2
 ~,
\end{eqnarray}
where $I_0 = (I-1) / 2$, and $I \le 2j$.

Next we discuss the case of $T=0$. Here ($T_2$,  $T'_2$) can take
following values: (1,1) and (0,0), and the corresponding
requirements for ($J,K$) are ($J={\rm even},K={\rm even}$) and
($J={\rm odd},K={\rm odd}$), respectively. Similarly we obtain the
following results
\begin{eqnarray}
& & 6 D_{I(T=0)} \nonumber \\
&=& \sum_{{\rm even} ~J ~ {\rm even }~K}  ( 1 + (-)^{I}
\delta_{JK} - 36 (2J+1) (2K+1)    \nonumber \\
&&  \times     \left\{
     \begin{array}{ccc}
     j    & j  & J \\
     j    & j  & K \\
     J    & K  & I   \end{array}
     \right\} ~
     \left\{
     \begin{array}{ccc}
     \frac{1}{2}    & \frac{1}{2}  & 1 \\
     \frac{1}{2}    & \frac{1}{2}  & 1 \\
     1    & 1 & 0   \end{array}
     \right\}   ~)~ \nonumber \\
&+& \sum_{{\rm odd} ~J ~ {\rm odd} ~K}  ( 1 + (-)^{I}
\delta_{JK} - 4 (2J+1) (2K+1)    \nonumber \\
&&  \times     \left\{
     \begin{array}{ccc}
     j    & j  & J \\
     j    & j  & K \\
     J    & K  & I   \end{array}
     \right\} ~
     \left\{
     \begin{array}{ccc}
     \frac{1}{2}    & \frac{1}{2}  & 0 \\
     \frac{1}{2}    & \frac{1}{2}  & 0 \\
     0    & 0 & 0   \end{array}
     \right\}   ~)~ \nonumber \\
& = & \sum_{{\rm even} ~J ~ {\rm even }~K}  ( 1 + (-)^{I}
\delta_{JK}) \nonumber \\
&+&  \sum_{{\rm odd} ~J ~ {\rm odd }~K}  ( 1 + (-)^{I}
\delta_{JK} ) - \frac{1}{2} S(j^4, {\rm odd} ~J ~ {\rm odd } ~K ~)   \nonumber \\
&& +\frac{1}{2} S(j^4, {\rm even} ~J ~ {\rm even }~K ~) ~.
\label{result2}
\end{eqnarray}
In derivation of Eq. (\ref{result2}) we used following relation:
\begin{eqnarray}
&&  \left\{   \begin{array}{ccc}
     \frac{1}{2}    & \frac{1}{2}   & 1 \\
     \frac{1}{2}    & \frac{1}{2}   & 1 \\
     1    & 1  & 0   \end{array}
     \right\}   = - \frac{1}{18} ~,
   \left\{  \begin{array}{ccc}
     \frac{1}{2}     & \frac{1}{2}   & 0 \\
     \frac{1}{2}    & \frac{1}{2}   & 0 \\
     0    & 0  & 0   \end{array}
     \right\}   =  1/2 ~.  \nonumber
\end{eqnarray}
To simplify Eq.(\ref{result2}), again one should consider the
number of combinations of $J$ and $K$, which are very complicated.
Below we show our final results.

When $I \ge 2j$ and $T=0$, we use Eq.(\ref{result2}), Eqs. (11)
and (19-20) of Ref. \cite{sum-rule}, and obtain
\begin{equation}
 D_{I(T=0)} = \left( \left[\frac{I_0}{3} \right] +1 \right)
 \left( \frac{3}{2} \left[\frac{I_0}{3} \right]+1 + (I_0 {\rm ~mod~ } 3)  \right)
 ~,
\end{equation}
where $I_0 = (I_{\rm max} -I)/2$ and $I$ is even. For $I$ is odd
and $I \ge 2j$, one has $D_{I(T=0)} = D_{(I+3)(T=0)}$.

When $I \le 2j$ and $T=0$, we use Eq. (\ref{result2}), Eqs.
(14-15) and (17-18) of Ref. \cite{sum-rule}. Let us define $I=6k +
\kappa$, $L=\left[ \frac{j-\frac{6k+3}{2}}{3} \right]$ and $m=((j
- (6k+3)/2) ~ {\rm mod ~} 3) \equiv (j-3/2) ~ {\rm mod ~} 3$. For
$\kappa = 0$, we obtain
\begin{eqnarray}
&& D_{I=6k(T=0)} = (2 + 6k) L
\nonumber \\
&&+ (2k+1)m + \frac{3}{2} k (k+3) + 1 ~,  \label{eq14}
\end{eqnarray}
and for $\kappa = 3$, we have
\begin{eqnarray}
&& D_{I=6k+3(T=0)} = (2 + 6k) L
\nonumber \\
&& + (2k+1)m + \frac{3}{2} k (k+1) ~.
\end{eqnarray}
We can see the following relation:
\begin{equation}
  D_{I=6k(T=0)} - D_{I=6k+3(T=0)} = 3k   + 1.
\end{equation}

For $\kappa = 1$, we have
\begin{eqnarray}
&& D_{I=6k+1(T=0)} = 2 k j  -  \frac{1}{2}k(9k+1)   ~,
\end{eqnarray}
Note $D_{I=1,T=0} = 0$. For $\kappa=4$ we obtain
\begin{equation}
D_{I=6k+4(T=0)} = 2 (k+1) j -  \frac{1}{2}(k+1)(9k+4)  ~.
\end{equation}
 We have the following relation:
\begin{equation}
 D_{I=6k+4(T=0)} - D_{I=6k+7(T=0)} = 3(k+1).
\end{equation}

For  $\kappa = 2$, we have
\begin{eqnarray}
&& D_{I=6k+2(T=0)} = (4 + 6k) L
\nonumber \\
&& + (2k+1)m  + \frac{1}{2} (k +1)(3k+4)
~, \nonumber \\
\end{eqnarray}
and for $\kappa = 5$, we have
\begin{eqnarray}
&& D_{I=6k+5(T=0)} = (4 + 6k) L
\nonumber \\
&& + (2k+1)m  + \frac{1}{2} k(3k+1) ~.
\end{eqnarray}
One can notice
\begin{equation}
  D_{I=6k+2(T=0)} - D_{I=6k+5(T=0)} = 3k+2.
\end{equation}

We similarly obtain $D_{I(T=\frac{1}{2})}$ for three nucleons:
\begin{eqnarray}
&& D_{I(T=1/2)}  \nonumber \\
&&= \sum_{ {\rm even~}J} \left(1 - 6 (2J+1)
 \left\{   \begin{array}{ccc}
     j    & j   & J \\
     j    & I   & J \\       \end{array}      \right\}
 \left\{   \begin{array}{ccc}
     1/2    & 1/2   & 1 \\
     1/2    & 1/2   & 1 \\       \end{array}      \right\} \right)
     \nonumber \\
&&+ \sum_{ {\rm odd~}J} \left(1 - 2 (2J+1)
 \left\{   \begin{array}{ccc}
     j    & j   & J \\
     j    & I   & J \\       \end{array}      \right\}
 \left\{   \begin{array}{ccc}
     1/2    & 1/2   & 0 \\
     1/2    & 1/2   & 0 \\       \end{array}      \right\} \right)
     \nonumber \\
&& =\sum_{ {\rm even~}J}  \left(1 -   (2J+1)
 \left\{   \begin{array}{ccc}
     j    & j   & J \\
     j    & I   & J \\       \end{array}      \right\}
 \right) \nonumber \\
&&+  \sum_{ {\rm odd~}J} \left(1 +   (2J+1)
 \left\{   \begin{array}{ccc}
     j    & j   & J \\
     j    & I   & J \\       \end{array}      \right\}
  \right) ~, \label{sum3}
\end{eqnarray}
where the following six-$j$ symbols are used:
\begin{eqnarray}
 && \left\{   \begin{array}{ccc}
     1/2    & 1/2   & 1 \\
     1/2    & 1/2   & 1 \\       \end{array}      \right\} =  \frac{1}{6},
 \left\{   \begin{array}{ccc}
     1/2    & 1/2   & 0 \\
     1/2    & 1/2   & 0 \\       \end{array}      \right\} =
   -  \frac{1}{2}
     ~.
\end{eqnarray}
By using Eq. (\ref{sum3}) and  the sum rules of six-$j$ symbols
(Eqs.(A3) and (A8) obtained in Ref. \cite{Zhao-prc70}), one can
easily obtain $D_{I(T=1/2)}$ for three nucleons.  For $I \le j$,
\begin{eqnarray}
& & D_{I(T=\frac{1}{2})} = 1 + 2\left[ \frac{( I-1/2 )}{3} \right]
\nonumber \\
&&  + \delta_{(( I-1/2 ) {\rm ~ mod ~}3), 2} ~, \label{f3t11}
\end{eqnarray}
and for $I \ge j$, we have
\begin{eqnarray}
& & D_{I(T=\frac{1}{2})} = 1 + \left[ \frac{( I_{\rm max} - I)}{3}
\right] ~, \label{f3t12}
\end{eqnarray}
where $I_{\rm max} = 3j -1 $.

According to Eqs.(1-2) of Ref. \cite{Zhao}, when $I \le j$,
\begin{eqnarray}
&& D_{I(T=\frac{3}{2})} = \left[ \frac{2I+3}{6} \right] ~;
\label{t=3/2-1}
\end{eqnarray}
and when $I \ge j$,
\begin{eqnarray}
&& D_{I(T=\frac{3}{2})} = \left[ \frac{3j-3 - I}{6} \right] +
\delta_I ~,  \label{t=3/2-2}
\end{eqnarray}
where
\begin{eqnarray}
 && \delta_I = \left\{   \begin{array}{ll}
  0   & {\rm if~} ((3j-3)-I) {~\rm mod~} 6 =1 ~, \\
  1   & {\rm otherwise}   ~. \end{array}      \right. \nonumber
\end{eqnarray}
Note that $j \le I \le 3j -3$ in Eq. (\ref{t=3/2-2}). Comparing
Eqs. (\ref{t=3/2-1}) and (\ref{t=3/2-2}) with Eqs. (\ref{f3t11})
 and (\ref{f3t12}), one easily sees that $D_{I(T=\frac{1}{2})} - 2
D_{I(T=\frac{3}{2})}$ has a a modular behavior. We obtain that
when $I\le j$
\begin{eqnarray}
&& D_{I(T=\frac{1}{2})}-2 D_{I(T=\frac{3}{2})} \nonumber \\
&&= \left\{
     \begin{array}{ll}
     -1    & {\rm if~} (I - 1/2){\rm ~ mod~} 3 = 1     \\
     1    & {\rm otherwise~}   \end{array}
     \right.  \nonumber \\
&&= \left\{
     \begin{array}{ll}
     -1    & {\rm if~} 2I {\rm ~ mod~} 3 = 0     \\
     1    & {\rm otherwise~}   \end{array}
     \right.  \nonumber \\
&& = 1-2 \delta_{2I {\rm ~mod ~} 3, 0} ~,
\end{eqnarray}
and when $I \ge j$
\begin{eqnarray}
&& D_{I(T=\frac{1}{2})}- D_{I(T=\frac{3}{2})} \nonumber \\
&&= \left\{
     \begin{array}{ll}
     1    & {\rm if~} (3j-1 -I){\rm ~ mod~} 6 = 0     \\
     1    & {\rm if~} (3j-1 -I){\rm ~ mod~} 6 = 1     \\
    -1    & {\rm if~} (3j-1 -I){\rm ~ mod~} 6 = 2     \\
     2    & {\rm if~} (3j-1 -I){\rm ~ mod~} 6 = 3     \\
     0    & {\rm if~} (3j-1 -I){\rm ~ mod~} 6 = 4     \\
     0    & {\rm if~} (3j-1 -I){\rm ~ mod~} 6 = 5. \end{array}
     \right. \nonumber \\
&& = \left( (\left[\frac{3j-1-I}{2} \right] + 1 ) {\rm ~ mod ~} 3
\right) \nonumber \\
&& + \left\{
     \begin{array}{ll}
     -3    & {\rm if~} (3j-1-I) {\rm ~ mod~} 6 = 2     \\
     0    & {\rm otherwise~} . \end{array}
     \right.
\end{eqnarray}

For $n=4$, a modular behavior for $D_{I(T=0)}- D_{I(T=2)}$ was
recently found in Ref. \cite{Zamick1} by Zamick and A. Escuderos,
who proved that
\begin{eqnarray}
&& D_{I (T=0)} -2 D_{I(T=2)}   \nonumber \\
&=& 2 \sum_{{\rm even} ~J ~ {\rm even ~K}}   (2J+1) (2K+1) \left\{
     \begin{array}{ccc}
     j    & j  & J \\
     j    & j  & K \\
     J    & K  & I   \end{array}
     \right\}  ~ . \nonumber \\
     \label{Zamick}
\end{eqnarray}
We can prove that our results are consistent with this relation.
Let us take $I=12k$ (but $I\le 2j-3$) as an example. We define
$L=\left[ \frac{j-\frac{12k+3}{2}}{3} \right]$ and $m= (j-3/2) ~
{\rm mod ~} 3$. By using Eqs. (\ref{eq5}) and (\ref{eq14}), we
obtain
\begin{eqnarray}
&& D_{I=0(T=0)} - 2 D_{IT=2} \nonumber \\
&& = (12k+2) L + (4k+1)m+3k(2k+3) + 1 \nonumber \\  && ~~ -
2\left[ (6k+1)L + 2km + 1 + 3k(k+1) \right] \nonumber \\  && = 3k+
m -1~ \nonumber  ~.
\end{eqnarray}
This is the right hand side of Eq. (\ref{Zamick}) for $I=12k$,
according to Eq. (14) of Ref. \cite{sum-rule}. The cases of other
$I$ values can be proved similarly.

\vspace{0.4in}

 To summarize, in this paper we obtained, for the
first time, algebraic formulas for the number of states with spin
$I$ and isospin $T$ for three and four nucleons in a single-$j$
shell by using the sum rules of six-$j$ and nine-$j$ symbols
obtained in Ref. \cite{sum-rule} for identical particles and the
well known sum rule for two-body coefficients of fractional
parentage.  We also showed that $D_{I(T=\frac{1}{2})}- 2
D_{I(T=\frac{3}{2})}$ has a simple modular behavior for $n=3$. We
note without   details that same procedures can be applied to
obtain number of states with given spin $I$ and $F$ spin for three
and four bosons in the interacting boson models.

\vspace{0.4in}

Acknowledgement:  One of the authors (YMZ) would like to thank the
National Natural Science Foundation of China for supporting this
work under Grant Nos. 10545001 and 10575070.


\end{document}